\documentclass[12pt,aps]{revtex4}

\input{epsf}
\begin{document}

\title{Spectrum and eigen functions of  the operator  $H_U f(x)\equiv f(U-1/x)/x^2$
 and  strange attractor's density for  the mapping  $x_{n+1}=1/(U-x_n)$}

\author{Kozlov G.G.}
\begin{abstract}
We solve exactly the spectral problem for the non-Hermitian operator  $H_U f(x)\equiv f(U-1/x)/x^2$.
  Despite the absence of orthogonality,  the eigen functions
 of this operator allow one to construct in a simple way the expansion of an arbitrary
 function in series. Explicit formulas for the expansion coefficients are
 presented.
 This problem is shown to be connected with  that  of calculating
  the strange attractor's
 density for the map  $x_{n+1}=1/(U-x_n)$.
  The explicit formula
 for  the strange attractor's
 density for this map  is derived.
  All results are confirmed  by direct computer simulations.
\end{abstract}
\maketitle 
\vskip5mm

The author of this note is not an expert in the theory of strange attractors.
 So, the results presented below may appear to be well-known to those who work
 in this field.
If the material of this note will be found familiar, totally or to some extend,
 for some of the readers, the author will much appreciate sending him
 appropriate references which will be included in to the next version of this
 publication.

Let us consider the linear operator $H_U$, depending on the
parameter  $U$ and acting on the arbitrary function  $f(x)$ (with
real $x$) in accordance with the following definition
\begin{equation}
H_U f(x)\equiv {1\over x^2}f\bigg(U-{1\over x}\bigg)
\end{equation}
 and consider the spectral problem for this  operator i.e. find its
 eigen numbers $\lambda_U$ and eigen functions $\sigma_U^\lambda(x)$:
  \begin{equation}
  H_U\sigma_U^\lambda(x)=\lambda_U\sigma_U^\lambda(x)
  \end{equation}

We solve this problem under the following conditions: (i)
 the eigen functions of (1) must be {\it  finite}, (ii)
 the parameter $U$ obeys the inequality $|U|<2$  which  corresponds to the regime of strange
 attractor for the mapping  $x\rightarrow U-1/x$ which is coinside with
   operator (1) up to the factor $1/x^2$  providing the equality
 of the integrals  $\int f(x)dx=\int H_Uf(x) dx$.

Using (2) and (1) we write (we omit indexes  $U$ and  $\lambda$):

\begin{equation}
\sigma(x)={1\over\lambda}{1\over x^2}\sigma(U-1/x)=
{1\over\lambda^2}\bigg({1\over x (U-1/x)}\bigg)^2\sigma(U-1/U-1/x)=
\end{equation}
$$
{1\over\lambda^3}\bigg({1\over x
(U-1/x)(U-1/U-1/x)}\bigg)^2\sigma(U-1/U-1/U-1/x)=
$$
$$
{1\over\lambda^n}\bigg(\overbrace{{1\over x
(U-1/x)(U-1/U-1/x)...}}^n\bigg)^2\sigma(\overbrace{U-1/U-1/U-1/...1/x)}^n
$$
Define $T$ as:
\begin{equation}
T x\equiv U-1/x
\end{equation}
Then one can rewrite (3) as:
\begin{equation}
\sigma(x)={1\over \lambda^n}\bigg[\prod_{k=0}^{n-1}T^k \hskip1mmx\bigg]^{-2}\sigma(T^n\hskip1mm x)
\end{equation}
Now note that the eigen function of (1) $H_U$ with the
 eigen number equal to unit $\lambda=1$
 can be written in the form:
 \begin{equation}
 \sigma_U^{\lambda=1}(x)={\cal L}_U(x)\equiv{\sqrt{4-U^2}\over 2\pi}{1\over x^2-Ux+1}
 \end{equation}
We normalize this function by the condition
 $\int {\cal L}_U(x)dx=1$.

Now one can express the complex fractions entering (3) as:

\begin{equation}
\bigg[\prod_{k=0}^{n-1}T^k \hskip1mmx\bigg]^{-2}={{\cal L}_U(x)\over{\cal L}_U(T^n\hskip1mm x)}
\end{equation}

 and consequently

 \begin{equation}
 {\sigma(x)\over\sigma(T^n\hskip1mm x)}={1\over\lambda^n}{{\cal L}_U(x)\over{\cal L}_U(T^n\hskip1mm x)}
 \end{equation}
This formula is valid for an arbitrary eigen function of (1) and
 for any integer $n$.
 Taking into account the finiteness of the eigen functions of (1)
 we come to the conclusion that
  $|\lambda|=1$,
Otherwise, the left-hand side of (8) either vanishes or diverges at
$n\rightarrow\infty$  which contradicts to the finiteness of the eigen functions
   $\sigma(x)$.
   Thus:
   \begin{equation}
   \lambda=\exp(\imath\phi)
   \end{equation}

with real $\phi$.
 Equalising the module of both parts of (8) we obtain:
 \begin{equation}
 |\sigma(x)|={\cal L}_U(x)
 \end{equation}
and consequently:
\begin{equation}
\sigma(x)={\cal L}_U(x)\exp(\imath\theta(x)),
\end{equation}
  with the function  $\theta(x)$ being real. Using (8) and (9) we obtain
 for  $\theta(x)$ the following equation:
  \begin{equation}
  \exp\imath[n\phi-\theta(T^n x)+\theta(x)]=1
  \end{equation}
   whence

\begin{equation}
n\phi-\theta(T^n x)+\theta(x)=2\pi \hbox{Mi}_n(x)
\end{equation}

where Mi$_n(x)$  is an arbitrary function
 taking the integer values.
 Suppose we find the function $\theta(x)$ obeying:
\begin{equation}
\theta(T\hskip1mm x)-\theta(x)=
\theta(U-1/x)-\theta(x)=\phi + 2\pi M(x)
\end{equation}
with $M(x)$  being an integer function of $x$.
 Then:
 \begin{equation}
\theta(T^{k+1}\hskip1mm x)-\theta(T^k x)=\phi+ 2\pi M(T^k x).
 \end{equation}
Now summing (15) over $k$ from zero to $n-1$ we obtain:

\begin{equation}
\theta(T^n x)-\theta(x)=n\phi+2\pi\sum_{k=0}^{n-1}M(T^k x).
\end{equation}
Here $T^0x=x$.
 Consequently the function $\theta(x)$ defined from (14) satisfy (13)
 with
 $$
 \hbox{Mi}_n(x)=-\sum_{k=0}^{n-1}M(T^k x)
 $$
 Thus the problem reduced to solving the equation (14).
We will see below that  solution of this equation can be found only for
 some particular value of angle $\phi$.

 To solve the equation (14) let us consider an auxiliary equation
\begin{equation}
{G(U-1/x)\over G(x)}=\xi
\end{equation}

Note that  $\ln G(x)$ obey the equation similar to
 (14). Construct the solution of (17) in the form:
 \begin{equation}
 G(x)={1+ax\over 1+bx}
 \end{equation}

Substituting (18) in to (17) one can obtain the set of equations for
 $a$, $b$ и $\xi$. The solution of this set of equations gives:
\begin{equation}
\xi={U^2-2\pm\imath|U|\sqrt{4-U^2}\over 2}
\end{equation}
\begin{equation}
a=-{1+\xi\over U}
\end{equation}

\begin{equation}
b=-{1+1/\xi\over U}
\end{equation}

If $|U|<2$ (the strange attractor regime
 for mapping (4))
 then $|\xi|=1$.
 The sign of square root in (19) is not important.
   Then:
 \begin{equation}
 \xi=\exp\imath\gamma
 \end{equation}
 with
 \begin{equation}
 \gamma=\hbox{arctg}\bigg(
{|U|\sqrt{4-U^2}\over U^2-2}\bigg)
 +\cases{0\hskip3mm\hbox{when}\hskip1mm \sqrt{2}<|U|<2\cr \pi
 \hskip3mm\hbox{when}\hskip1mm \sqrt{2}>|U|}
 \end{equation}
 Here we take into account the continuity of $\xi(U)$ function (19).
 Then it is easy to see that
 $|G(x)|=1$ and consequently:
\begin{equation}
G(x)=\exp\imath\ae
\end{equation}
 with
 \begin{equation}
 \hbox{tg}\hskip1mm{\ae\over 2}={x\sin\gamma\over x(1+\cos\gamma)-U}
 \end{equation}
 and consequently:
\begin{equation}
\ae=2\hskip1mm\hbox{arctg}\bigg[
{x\sin\gamma\over x(1+\cos\gamma)-U}
\bigg]+2\pi M(x)
\end{equation}
where $M(x)$ is some integer function of $x$.
 Then one can see that
\begin{equation}
\hbox{Im}\ln G(x)=\ae(x)
\end{equation}
 satisfy the equation
 (14) with  $\phi=\gamma$.
 Thus the final solution of equation (14) can be written as:
\begin{equation}
\theta(x)=2\hskip1mm\hbox{arctg}\bigg[
{x\sin\phi\over x(1+\cos\phi)-U}
\bigg]
\end{equation}
\begin{equation}
 \phi=\hbox{arctg}\bigg(
{|U|\sqrt{4-U^2}\over U^2-2}\bigg)
 +\cases{0\hskip3mm\hbox{when}\hskip1mm \sqrt{2}<|U|<2\cr \pi
 \hskip3mm\hbox{when}\hskip1mm \sqrt{2}>|U|}
 \end{equation}
and
\begin{equation}
\theta(U-1/x)-\theta(x)=\phi + 2\pi M(x)
\end{equation}

 Now one can see that the function
 $\theta_n(x)\equiv n\hskip1mm\theta(x)$ where $n$ is integer,
 satisfy the equation
  (14) with $\phi\rightarrow\phi_n=n\phi$.

Thus one can write the set of eigen functions  $\sigma^n_U(x)$
 and eigen numbers  $\lambda_n$ of (1) as:
\begin{equation}
\sigma^n_U(x)={\cal L}_U(x)\exp\imath n\theta(x),\hskip10mm\lambda_n=\exp\imath
n\phi
\end{equation}
where $n$ is integer, and ${\cal L}_U(x)$, $\theta(x)$ and $\phi$
 are defined by (6), (28) and (29).

\section{Properties of the eigen functions and the relevant expansion}

Functions (31) should possess the following property:
 each of these functions (except for ${\cal L}_U(x)$ with
$\lambda=1$) has zero integral.  To show this let us integrate over $x$ the
equation for an arbitrary function

\begin{equation}
\int{dx\over x^2}\sigma(U-1/x)=\lambda\int\sigma(x)dx.
\end{equation}
  Replacing  $1/x\rightarrow x$ in the left-hand side  we obtain
 \begin{equation}
 (1-\lambda)\int\sigma(x)dx=0
 \end{equation}
 Thus if  $\lambda\ne 1$, then
 \begin{equation}
 \int\sigma^{\lambda\ne 1}(x)dx=0
 \end{equation}

The next property  of functions (31) which we use below is:

\begin{equation}
\sigma_U^n(x)\exp\imath m\theta_U(x)=\sigma_U^{n+m}(x)
\end{equation}

These properties allows one to expand an arbitrary function $f(x)$ in series
 using the set of functions (31) as follows. Suppose that  function $f(x)$
 can be expanded in series
\begin{equation}
f(x)=\sum_{n=-\infty}^{+\infty} C_n \sigma_U^n(x)=
\sum_{n=-\infty}^{+\infty} C_n {\cal L}_U(x)\exp[\imath n\theta_U(x)].
\end{equation}

After multiplying both sides of (36) by
 $\exp-\imath m \theta_U(x)$ and integrating, one can
 obtain the following expression for the coefficients
 $C_m$:
\begin{equation}
C_m={\int f(x)\exp\big[-\imath m\theta_U(x)\big]dx\over \int{\cal L}_U(x)dx}=
\int f(x)\exp\big[-\imath m\theta_U(x)\big]dx
\end{equation}

The function
 $\theta_U(x)$ is discontinuous at
$$
x_0={U\over 1+\cos\phi}
$$
It can be converted into continuous one by adding  proper function
 in the form $2\pi M(x)$
(with $M(x)$ being integer function).
 The function
  $\theta_U(x)$  defined in this way will be as follows:
\begin{equation}
\theta_U(x)=2\hskip1mm\hbox{arctg}\bigg[
{x\sin\phi\over x(1+\cos\phi)-U}
\bigg]+\cases{0\hskip20mm\hbox{when}\hskip2mm
x<x_0\cr -2\pi |U|/U\hskip5mm\hbox{when}\hskip2mm x>x_0}
\end{equation}

The possibility of expansion (36), (37) was verified by
 direct computer calculations for various functions $f(x)$.

By simple algebra  functions (31) $\sigma_U^n(x)$ can
 be rewritten as:
\begin{equation}
\sigma_U^n(x)={\cal L}_U(x)\bigg({x(|U|+\imath\sqrt{4-U^2})-2U/|U|
\over x(|U|-\imath\sqrt{4-U^2})-2U/|U|}\bigg)^n
\end{equation}
$$
{\cal L}_U(x)={\sqrt{4-U^2}\over 2\pi}{1\over x^2-Ux+1}
$$
$$
\exp[\imath\theta_U(x)]={x(|U|+\imath\sqrt{4-U^2})-2U/|U|
\over x(|U|-\imath\sqrt{4-U^2})-2U/|U|}
$$
\newpage
 Omitting some unimportant coefficients one can finally present
 the eigen functions of (1) in the form:
\begin{equation}
\sigma_U^n(x)={1\over 2\pi\imath}{|U|\over U}\bigg({1\over x-R_U}-{1\over
x-R^*_U}\bigg)\bigg({x-R_U^*\over x-R_U}\bigg)^n
\end{equation}
\begin{equation}
R_U\equiv{U^2+\imath|U|\sqrt{4-U^2}\over 2U}
\end{equation}
\begin{equation}
R_U^*\equiv{U^2-\imath|U|\sqrt{4-U^2}\over 2U}
\end{equation}
\begin{equation}
R_U^*R_U=1
\end{equation}

An expansion of an arbitrary function by functions (40) has the form:

\begin{equation}
f(x)=\sum_{n=-\infty}^{+\infty} C_n \sigma_U^n(x)
\end{equation}
где
\begin{equation}
C_n=\int f(x)\bigg({x-R_U\over x-R_U^*}\bigg)^n dx
\end{equation}

The example of above expansion for various number of "harmonics" is presented on
Fig.1.

\section{
Linear-fraction mapping, an explicit formula for an arbitrary number of steps, strange
attractor, cycles}

Consider an arbitrary linear-fractional mapping of the form:

\begin{equation}
x_{n+1}={1+ax_n\over b+c x_n}.
\end{equation}

One can  reduce (46) to the form:
 \begin{equation}
 x_{n+1}={1\over U-x_n}
 \end{equation}
 by replacing $ x_n\rightarrow k_1 x_n+k_2$ with appropriate constants  $k_1$ и
 $k_2$. For this reason without loss of generality one can study mapping (47).
 Introducing new variable
  $y_n$  as

\begin{equation}
\overbrace{
x_n={1+y_n\over \theta+\xi y_n}
}^{O_{y\rightarrow x}}
\end{equation}
and back
\begin{equation}
\overbrace{
y_n={1-\theta x_n\over \xi x_n-1}
}^{O_{x\rightarrow y}}
\end{equation}

one can reduce the corresponding mapping for variable $y_n$ to the simple form:

\begin{equation}
y_{n+1}=\ae y_n
\end{equation}
 with the following values of parameters entering (48),(49) and (50):
 \begin{equation}
 \xi={1\over 2}\bigg[
U-\sqrt{U^2-4}
\bigg]
 \end{equation}
$$
\theta={1\over 2}\bigg[
U+\sqrt{U^2-4}
\bigg]
$$
$$
\ae={U-\sqrt{U^2-4}\over U+\sqrt{U^2-4}}
$$

or

 \begin{equation}
 \xi={1\over 2}\bigg[
U+\sqrt{U^2-4}
\bigg]
 \end{equation}
$$
\theta={1\over 2}\bigg[
U-\sqrt{U^2-4}
\bigg]
$$
$$
\ae={U+\sqrt{U^2-4}\over U-\sqrt{U^2-4}}
$$

Thus if we perform  mapping (47) $n$ times using $x_0$ as a starting point we
 obtain for $x_n$ the following expression:

 \begin{equation}
 x_n={1+\ae^ny_0\over\theta+\xi\ae^n y_0}\hskip5mm\hbox{where}\hskip3mm
 y_0={1-\theta x_0\over \xi x_0-1}
 \end{equation}

In the case of strange attractor regime (i.e. when $U^2-4<0$ )
  $\ae=\exp\imath\phi$ with
$\phi=2$arctg$[\sqrt{4-U^2}/U]$.
If $\phi=2\pi/n$, with $n$ integer, then $x_{m+n}=x_m$
 and we obtain the simple cyclic regime. For given $n$ one can find the value
 of $U$ for which the simple cyclic regime is appeared:
\begin{equation}
U^2={4\over\hbox{tg}^2(\pi/n)+1}
\end{equation}
The general cyclic regime corresponds to the condition:
 $n\phi=2\pi m$ where $m$ and $n$ are integer.

\section{Density of strange attractor for mapping (47)}

The values $x_n$ obtained in the process of  successive  mapping (47) are aperiodically
 walking along the axis of numbers (when $|U|<2$, strange attractor regime).
 Let us define the relative density of  values $x_n$ after $N$ steps as
 $$
\rho(z)={1\over N}\sum_{n=0}^N\delta(z-x_n).
 $$
 We are interested in the ultimate density function defined as
\begin{equation}
\rho(z)\equiv\lim_{n\rightarrow\infty}{1\over N}\sum_{n=0}^N\delta(z-x_n).
\end{equation}

For this function one can obtain the functional equation as follows.
Using the fact that all quantities $x_n$ are successively derived   from each
other by mapping (47) we can write:
  \begin{equation}
  \rho(z)=\lim_{n\rightarrow\infty}{1\over N}\sum_{n=0}^{N-1}\delta\bigg(z-{1\over U-x_n}\bigg)
  +{1\over N}\delta(z-x_0).
  \end{equation}
Taking into account that
\begin{equation}
\delta\bigg(z-{1\over U-x}\bigg)={1\over z^2}\delta\bigg(U-x-{1\over z}\bigg)
\end{equation}
 we obtain:
 \begin{equation}
\rho(z)=\lim_{n\rightarrow\infty}{1\over N}\sum_{n=0}^N{1\over z^2}\delta\bigg(U-{1\over z}-x_n\bigg)
  +{1\over N}\delta(z-x_0)-{1\over N}\delta(U-1/z-x_N).
 \end{equation}
 Now one can see that at $N\rightarrow\infty$:
 \begin{equation}
 \rho(z)={1\over z^2}\rho(U-1/z).
 \end{equation}
 We see that the density we are interested in is the eigen function of the operator
  $H_U$, with eigen value equal to unit and therefore has the form of  lorentzian (6) :
 \begin{equation}
\rho(z)={\cal L}_U(z)={\sqrt{4-U^2}\over 2\pi}{1\over z^2-Uz+1}
 \end{equation}

Fig.2 shows the density of strange attractor obtained by computer simulation
 for various values of parameter $U$. The smooth curves were calculated by (60).

 All the aforesaid is valid for the case of the strange attractor regime
  $|U|<2$, when the density is not trivial.
   When $|U|>2$, the density has a delta-singularity
   at  $z=1/\theta$ for $|\ae|<1$ and $z=1/\xi$ for $|\ae|>1$.
Equation (59) can be generalized for the case of an arbitrary mapping
 of the form
$x_{n+1}=f(x_n)$.
Denoting the inverse mapping by
 $F$ (i.e.$x_n=F(x_{n+1})$, we obtain
 that the density of attractor of mapping
   $f$ satisfies the equation:
\begin{equation}
\rho(z)=F'(z)\rho\bigg(F(z)\bigg)
\end{equation}

\newpage
\section*{Captions}

Fig.1 The example of expansion in to series using the set of eigen functions of
non-hermitian operator (1)

Fig.2 The density of strange attractor obtained by computer simulation
 for various values of parameter $U$. The smooth curves were calculated by (60).


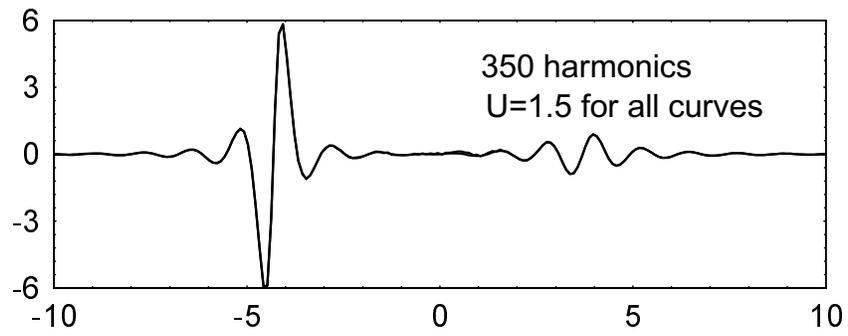
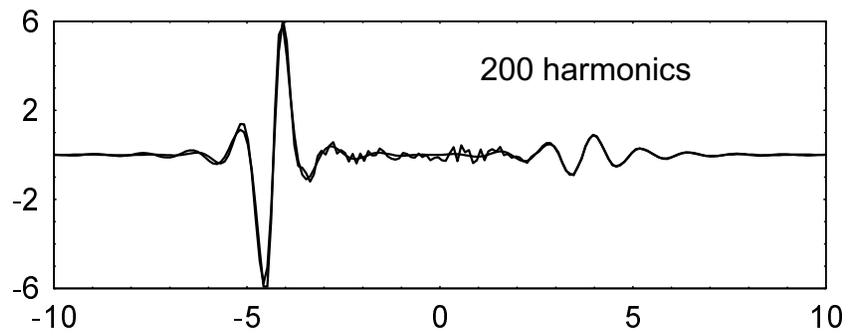
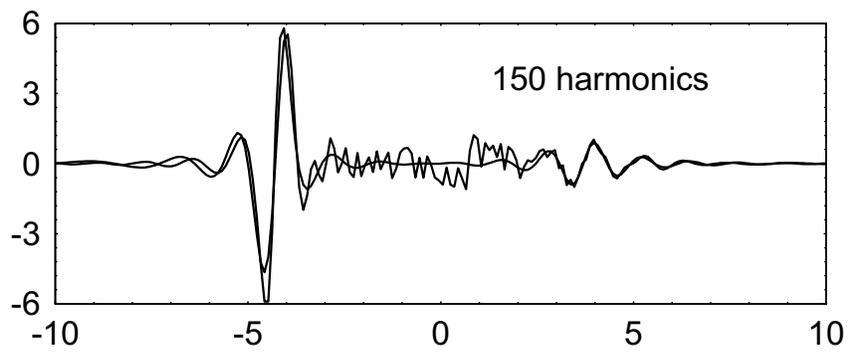
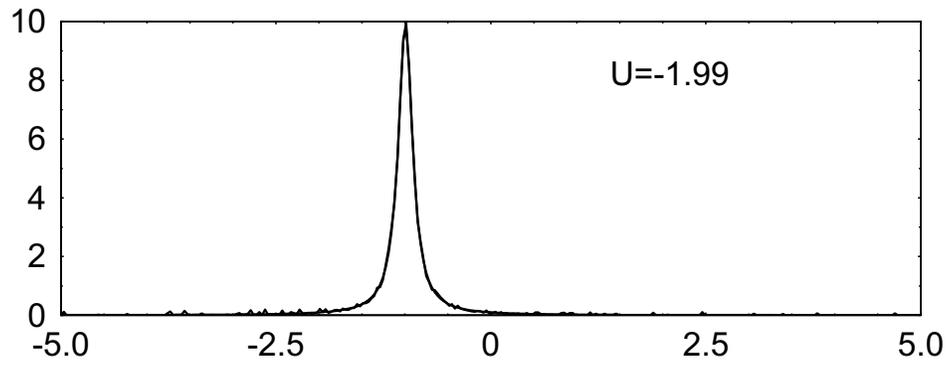
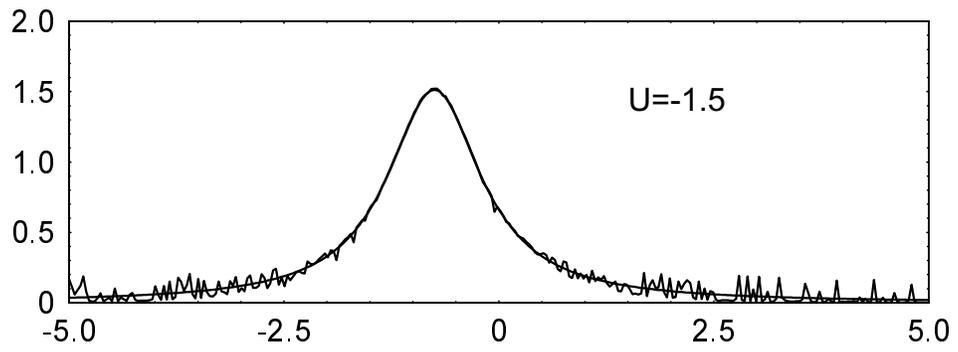
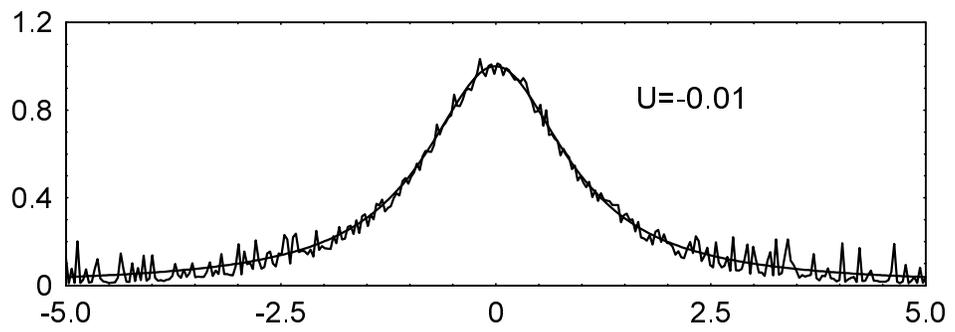
\end{document}